\documentclass[aps,prb,twocolumn,amsmath,amssymb,groupedaddress]{revtex4-2}
\usepackage[utf8]{inputenc}   
\usepackage[T1]{fontenc}      
\usepackage[english]{babel}   
\usepackage{lmodern}          
\usepackage{csquotes}         
\usepackage{amsmath}          
\usepackage{mathtools}        
\usepackage{amstext}          
\usepackage{amssymb}          
\usepackage{subdepth}         
\usepackage{xcolor}           
\usepackage[
  separate-uncertainty=true,
  per-mode=reciprocal,
  binary-units=true,
  range-phrase=--]{siunitx}   
\usepackage{braket}           
\usepackage{graphicx}         
\usepackage[pass]{geometry}   
\usepackage[pdftex]{hyperref} 
\usepackage[capitalize]{cleveref}

\DeclareMathOperator{\tr}{Tr}
\DeclareMathOperator{\im}{Im}
\DeclareMathOperator{\re}{Re}
\DeclareMathOperator{\expect}{E}
\DeclareMathOperator{\bias}{bias}
\DeclareMathOperator{\corr}{corr}
\DeclareMathOperator{\ee}{\mathrm e}

\newcommand{\cee}{\hat{c}^{\vphantom{\dagger}}}
\newcommand{\eff}{\hat{f}}
\newcommand{\cdag}{\hat{c}^\dagger}
\newcommand{\fdag}{\hat{f}^\dagger}

\newcommand{\order}{O}
\renewcommand{\i}{\mathrm{i}}
\newcommand{\Def}{\coloneqq}

\newcommand{\del}[1]{\partial_{#1}}
\newcommand*{\dif}{\mathop{}\!\mathrm{d}}
\newcommand*{\Dif}[1]{\mathop{}\!\mathrm{d^{#1}}}
\newcommand{\vect}[1]{\mathbf{#1}}
\newcommand{\mean}[1]{\bar{#1}}
\newcommand{\estim}[1]{\hat{#1}}
\newcommand{\wideestim}[1]{\widehat{#1}}

\newcommand{\dga}{D$\Gamma$A}
\newcommand{\bfdga}{\textbf{D$\mathbf{\Gamma}$A}}

\makeatletter
\newcommand{\thefontsize}{\f@size pt\par}
\makeatother

\allowdisplaybreaks[1]    

\hypersetup{
  pdfauthor = {Patrick Kappl, Markus Wallerberger, Josef Kaufmann, Matthias Pickem,
  Karsten Held},
  pdftitle = {Statistical error estimates in dynamical mean-field theory and extensions
  thereof},
  pdfsubject = {Jackknife Paper},
  pdfkeywords = {jackknife, DMFT, DGA},
  pdfdisplaydoctitle = true
}

\begin{document}
\title{Statistical error estimates in dynamical mean-field theory and extensions thereof}
\author{Patrick Kappl, Markus Wallerberger, Josef Kaufmann, Matthias Pickem, Karsten Held}
\affiliation{Institute of Solid State Physics, TU Wien, 1040 Vienna, Austria}
\date{\today}

\begin{abstract}
  We employ the jackknife algorithm to analyze the propagation of the statistical quantum
  Monte Carlo error through the Bethe--Salpeter equation. This allows us to estimate the
  error of dynamical mean-field theory calculations of the susceptibility and of dynamical
  vertex approximation calculations of the self-energy. We find that the different
  frequency components of the susceptibility are uncorrelated, whereas those of the
  self-energy are correlated. For improving the quality of the correlation matrix taking
  sufficiently many jackknife bins is key, while for reducing the standard error of the
  mean sufficiently many Monte Carlo measurements are necessary. We furthermore show that
  even in the case of the self-energy, the finite covariance does not have a sizable
  influence on the analytic continuation.
\end{abstract}

\maketitle

\section{Introduction}
\label{sec:introduction}
  Developing reliable theories for strong electronic correlation has proved a Herculean
  task. Three decades after its invention\cite{Metzner1989,Georges1992a,Jarrell1992},
  dynamical mean-field theory (DMFT) has become state-of-the-art to calculate strongly
  correlated models\cite{Georges1996} and
  materials\cite{Anisimov1997,Lichtenstein1998,Held2006,Kotliar2006,Held2007}.
  Notwithstanding, one of the core scientific tasks, namely providing a proper error
  estimate for such calculations, is still in its infancy. Error estimates which directly
  follow from the quantum Monte Carlo (QMC) simulation of the (self-consistently
  determined) DMFT impurity problem, e.g. for the magnetization or compressibility, have
  been provided already from the beginning of DMFT, see, e.g.,
  Refs.~\onlinecite{Jarrell1992,Ulmke1995,Held1997a}. But as the focus of such
  calculations is nowadays more on the one-particle spectral function or two-particle
  susceptibility, error estimates are by and large missing.

  This is because such an error estimate is nontrivial. First, calculating the spectrum or
  susceptibility requires complex, non-linear post-processing routines such as a maximum
  entropy analytic continuation or the Bethe--Salpeter equation, respectively. Besides
  proper error propagation through these post-processing procedures, there is, secondly,
  the iteration error, i.e., the difference between the numerical solution and the exact
  (true) DMFT fixed point. Third, DMFT is an approximation to the correlation problem
  itself, introducing a systematic error for finite dimensional systems.

  The latter, i.e., the error of DMFT as an approximation, can be systematically improved
  upon by either cluster\cite{Hettler2000} or diagrammatic extensions\cite{RMPVertex}.
  Cluster extensions replace the single impurity site with a small cluster, thus
  interpolating between DMFT and the full lattice problem. Given proper finite size
  scaling, enlarging the cluster then yields an error estimate. The huge numerical effort
  essentially restricts such cluster extensions however to small clusters on one- and
  two-dimensional lattices.

  Diagrammatic extensions, on the other hand, augment DMFT with a specific set of
  non-local Feynman diagrams. The dynamical vertex approximation
  (\dga)\cite{Toschi2007,Katanin2009}, a prominent diagrammatic extension, generalizes the
  concept of a local one-particle vertex (self-energy) in DMFT systematically to the two-,
  three-, $n$-particle vertex. For $n\rightarrow\infty$ one recovers the full problem. The
  corrections on the two-particle vertex level provide an error estimate for the DMFT
  approximation, and similarly those of the three-particle level as an error estimate for
  the \dga\ results on the two-particle vertex level. One can proceed in a similar fashion
  of approximating the error\cite{Ribic2017b} in the dual Fermion
  approach\cite{Rubtsov2008}. Since both the size of the vertices and the effort of the
  associated diagrammatic equations grow strongly in $n$, one is restricted to small $n$.

  In this paper, we focus on the error propagation through the post-processing procedure,
  specifically, the Bethe--Salpeter equation. This does not only involve the DMFT
  calculation of the susceptibility but also the \dga\ calculation of the self-energy
  which employs the same Bethe--Salpeter equation, and on top of that the Schwinger--Dyson
  equation. Input for both equations is the local two-particle Green's function and the
  properly converged DMFT one-particle Green's function. For multi-orbital systems or at
  low temperatures the two-particle Green's function is only accessible using QMC
  techniques, but requires quite some effort to calculate and hence has a substantial
  statistical QMC error. We employ the jackknife method
  \cite{miller-trustworthy-jackknife,miller-jackknife-review} to analyze the propagation
  of the QMC error through the Bethe--Salpeter equation and to estimate the error of the
  final DMFT susceptibility and \dga\ self-energy. Finally, we perform maximum entropy
  analytic continuations of the \dga\ self-energy, using the jackknife estimates for error
  and covariance.

  The remainder of the paper is organized as follows: \cref{sec:methods} describes the
  methods employed: DMFT, continuous-time quantum Monte Carlo, Abinitio\dga, and jackknife
  resampling. The input to our calculations, the two-particle Green's function, is
  analyzed in \cref{sec:2-particle-greens-function}. The main results for the \dga\
  self-energy and the DMFT susceptibilities are presented in
  \cref{sec:self-energy,sec:susceptibilities} respectively, while an analytical
  continuation of the \dga\ self-energy is done in \cref{sec:analytic-continuation}. A
  discussion and conclusion can be found in \cref{sec:conclusion}.

\section{Methods}
\label{sec:methods}
  In the following, we briefly recapitulate the essential steps for calculating the DMFT
  susceptibility and \dga\ self-energy, starting from the QMC calculation of the one- and
  two-particle Green's function. We restrict ourselves to the essential equations without
  discussing technical details such as reformulations of the equations or numerical
  efficiency. For these details we refer the reader to Ref.~\onlinecite{w2dynamics} as
  regards the QMC calculation with the w2dynamics package \cite{w2dynamics} and to
  Ref.~\onlinecite{Gull2011a} for a general review, to
  Refs.~\onlinecite{Galler2016,adga-cpc} as regards the calculation of the DMFT
  susceptibility and \dga\ self-energy with the {\em ab initio dynamical vertex
  approximation} (ADGA) package, and to Ref.~\onlinecite{RMPVertex} for a review. Further,
  we discuss the essential idea of the jackknife algorithm, again referring the reader to
  the review Ref.~\onlinecite{miller-jackknife-review} for a more detailed presentation.

  \subsection{Dynamical mean-field theory}
  \label{ssec:DMFTimp}
    In DMFT, we map the lattice model
    \begin{equation}
      H_{\text{lattice}} = U \sum_i \cdag_{i\uparrow} \cdag_{i\downarrow}
          \cee_{i\downarrow} \cee_{i\uparrow}
        + \sum_{\sigma,i,j} t_{ij}^{\vphantom{\dagger}}
          \cdag_{i\sigma}\cee_{j\sigma},
      \label{equ:lat}
    \end{equation}
    where $\cee_{i\sigma}$ annihilates a fermion of spin $\sigma$ on site $i$, $U$ is the
    on-site interaction, and $t_{ij}$ is the hopping matrix, onto an Anderson impurity
    model (AIM)\cite{Georges1992a,Jarrell1992}:
    \begin{align}
      H_{\text{AIM}} &= U \cdag_\uparrow \cdag_\downarrow
      \cee_\downarrow \cee_\uparrow \nonumber
      + \sum_\sigma \tilde{\epsilon}_\sigma \cdag_\sigma \cee_\sigma\notag\\
      &+ \underbrace{\sum_{\sigma,p} (V_{p} \cdag_\sigma \eff_{\sigma p}
        + V^*_{p} \fdag_{\sigma p} \cee_\sigma)}_{H_{\text{hyb}}}
      + \sum_p \epsilon_p \fdag_{\sigma p} \eff_{\sigma p} \;.
      \label{equ:aim}
    \end{align}
    Here $U$ is the impurity interaction which is the same as that of the original lattice
    problem; $V_{p}$ denotes the hybridization between the impurity (denoted by
    $\cdag_\sigma$ and $\cee_\sigma$ creation and annihilation operators for spin $\sigma
    \in \{\uparrow,\downarrow\}$) and bath site $p$ (denoted by corresponding
    $\fdag_{\sigma p}$ and $\eff_{\sigma p}$ operators) at energy $\epsilon_p$. In
    essence, DMFT determines an AIM that gives the same local one-particle physics as the
    lattice model where the corresponding parameters $V_{p}$ and $\epsilon_p$ (or the
    hybridization function $\Delta$) have to be determined
    self-consistently\cite{Georges1992a,Jarrell1992}. In what follows we assume that this
    DMFT self-consistency has been achieved to high accuracy. For the sake of simplicity,
    we have restricted the equations to the one-orbital problem, but the generalization to
    multi-orbital models is straightforward.
  
  \subsection{Continuous-time quantum Monte Carlo}
  \label{ssec:quantum-monte-carlo}
    In order to obtain the one- and two-particle Green's function for the Anderson
    impurity model \eqref{equ:aim}, we use continuous-time quantum Monte Carlo in the
    hybridization expansion (CT-HYB)\cite{Werner2006,Gull2011a} with worm
    sampling\cite{Gunacker15} as implemented in the w2dynamics
    package\cite{w2dynamics,Parragh2012}. CT-HYB with worm sampling proceeds in a
    three-step fashion: First, one splits the Hamiltonian $H$ into an interacting part,
    taken to be $H_I = H_{\text{hyb}}$, and the rest, $H_0$. Second, one expands both
    the partition function and the expectation value of some observable (\enquote{worm}
    $\mathcal{W}$) into a Dyson series with respect to $H_I$ and uses Wick's theorem to
    group diagrams into determinants. For the partition functions, this yields:
    \begin{equation}
    \begin{split}
      \mathcal Z &= \sum_{n=0}^\infty \frac{(-1)^n}{n!} \sum_{\sigma_1,\sigma'_1}
      \cdots\!\!\sum_{\sigma_n,\sigma'_n}
      \int_0^\beta \Dif{n}\tau\Dif{n}\tau'\\
        &\times\,\tr\!\left[T_\tau \ee^{-\beta H_\text{loc}}
        \prod_{i=1}^n \cdag_{\sigma_i}\!(\tau_i)\,\hat c_{\sigma'_i}\!(\tau'_i) \right]
                                \det\mathbf{\Delta},
      \label{equ:cthyb-z}
    \end{split}
    \end{equation}
    where $T_\tau$ denotes time ordering. The elements of the hybridization matrix
    $\mathbf{\Delta}$ are given by $\mathbf{\Delta}_{ij} = \Delta_{\sigma_i\sigma_j'}
    (\tau_{\vphantom{\sigma'_j}i} - \tau_{\vphantom{\sigma'_j}j}')$ with the hybridization
    function $\Delta_{\sigma\sigma'}(\tau) = \delta_{\sigma\sigma'} \sum_p
    V_{\vphantom{'}p} (\del{\vphantom{'}\tau} - \epsilon_{\vphantom{'}p})^{-1}
    V^*_{\vphantom{'}p}$.

    Similarly, we write down the hybridization expansion for the \emph{worm} operator
    $\mathcal{W(\{\tilde{\tau}\})}$. It can consist of several creation and annihilation
    operators with various number of time arguments. Most important examples are the one-
    and two-particle Green's function, where $\mathcal{W(\{\tilde{\tau}\})}$ stands for
    $T_\tau\cee_\sigma(\tau)\cdag_\sigma(\tau')$ and
    $T_\tau\cee_\sigma(\tau_1)\cdag_\sigma(\tau_2)
    \cee_\sigma(\tau_3)\cdag_\sigma(\tau_4)$, respectively. Other worm operators have been
    introduced in Refs.~\onlinecite{Gunacker2016,Kaufmann2017,Kaufmann2019}. For the
    sampling space of $\mathcal{W(\{\tilde{\tau}\})}$, we thus get:
    \begin{equation}
    \begin{split}
      \mathcal{Z}_\mathcal{W} &= \sum_{n=0}^\infty \frac{(-1)^n}{n!}
        \sum_{\sigma_1,\sigma'_1} \cdots\!\! \sum_{\sigma_n,\sigma'_n}
        \int_0^\beta \Dif{n}\tau\Dif{n}\tau'\dif\{\tilde{\tau}\}\\
      &\times\,\tr\!\left[T_\tau \ee^{-\beta H_\text{loc}}
        \mathcal{W}(\{\tilde{\tau}\})
        \prod_{i=1}^{n} \cdag_{\vphantom{'}\sigma_i}\!(\tau_i)\,\cee_{\sigma'_i}(\tau'_i)
        \right]
        \det\mathbf\Delta.
      \label{equ:cthyb-g}
    \end{split}
    \end{equation}
    Third, we combine both sampling spaces by taking the abstract sum $\mathcal{Z} +
    \eta\mathcal{Z}_\mathcal{W}$, where $\eta$ is a balancing parameter. The resulting
    space is sampled using Markov chain Monte Carlo.

    An estimator for the worm operator $\mathcal{W(\{\tilde{\tau}\})}$ is then simply
    given by:
    \begin{equation}
      \braket{\mathcal{W}_{}(\{\tilde{\tau}\})} = \frac{\mathcal{Z}_\mathcal{W}}
      {\mathcal{Z}}\braket{\sigma(\{\tilde{\tau}\})},
      \label{equ:1p-greens-function}
    \end{equation}
    where $\sigma(\{\tau_i\})$ is the indicator function of a configuration in
    $\mathcal{Z_W}$ with the matching times, $\mathcal{Z_W/Z}$ is the ratio of volumes
    between the two spaces. Let us note that in the case of the one- or two-particle
    Green's function one worm measurement is computationally cheaper than one measurement
    of the removal estimator in $Z$-sampling, but it also yields less information.

  \subsection{DMFT susceptibility}
  \label{ssec:DMFTsusc}
    In the following, we will make the transition from imaginary time to Matsubara
    frequencies, where the one-particle Green's function is
    \begin{equation}
      \label{equ:1pgf-mats}
      G_\sigma^\nu =  \int_0^\beta \dif\tau  e^{i\nu(\tau)}
      \braket{T_\tau \cee_\sigma(\tau) \cdag_\sigma(0)}
    \end{equation}
    and the two-particle Green's function is
    \begin{align}
      \label{equ:2pgf-mats}
      G_{\sigma \sigma'}^{(2),\nu\nu'\omega} = \frac{1}{\beta}
      \int_0^\beta &\dif\tau_1 \dif\tau_2 \dif\tau_3 
      e^{i[\nu\tau_1\! -\! (\!\nu\!+\!\omega)\tau_2 \!+\! (\!\nu'\!\!+\!\omega)\tau_3]}
      \notag\\
      &\braket{T_\tau \cee_{\vphantom{'}\sigma}(\tau_1) \cdag_{\vphantom{'}\sigma}(\tau_2)
        \cee_{\sigma'}(\tau_3) \cdag_{\sigma'}(0)}.
    \end{align}
    Note that here and in the following we use the notation of the w2dynamics code
    \cite{w2dynamics} which has the extra $1/\beta$ factor in the definition of the
    two-particle Green's function (\ref{equ:2pgf-mats}) compared to
    Ref.~\onlinecite{RMPVertex}. From these Green's functions, we can calculate the
    generalized local susceptibility
    \begin{equation}
      \label{equ:locsusc}
      \chi_{\sigma \sigma'}^{\vphantom{()}\nu\nu'\omega}=
      \beta G_{\sigma\sigma'}^{(2),\nu\nu'\omega}
      -\beta G^{\vphantom{()}}_{\nu\vphantom{\nu'}} G^{\vphantom{()}}_{\nu'}
      \delta^{\vphantom{()}}_{\vphantom{\nu'}\omega 0}.
    \end{equation}
    Diagrammatically, the above equation means that we remove one disconnected
    contribution from $G^{(2)}$. In the presence of SU(2) symmetry, i.e., in the
    paramagnetic phase, one can further restrict oneself to the following two spin
    combinations, also referred to as density ($d$) and magnetic ($m$) channel.
    \begin{equation}
      \label{equ:defgensus}
      \chi_{d/m,}^{\nu\nu'\omega}=\chi_{\uparrow\uparrow}^{\nu\nu'\omega}\pm\chi_{\uparrow
      \downarrow}^{\nu\nu'\omega}\; .
    \end{equation}

    From the local susceptibility $\chi_{r}^{\nu\nu'\omega}$ with $r\in\{d,m\}$ and the
    local bare bubble susceptibility $\chi_{\vphantom{\nu'}0}^{\nu\nu'\omega} = -\beta
    G_{\vphantom{\nu'}\nu} G_{\vphantom{\nu'}(\nu+\omega)} \delta_{\nu\nu'}$, we can
    determine the local irreducible vertex through an inversion of the local
    Bethe--Salpeter equation
    \begin{equation}
      \label{equ:defgamma}
      \Gamma_r=\chi_{r}^{-1} - \chi_{0}^{-1} \;.
    \end{equation}
    This equation is diagonal with respect to each bosonic frequency $\omega$ while the
    inversion itself is performed in the implicit matrix notation $\nu$ and $\nu'$. From
    $\Gamma$ and the local particle-hole bubble
    $\chi_{0,\vect{k}\vect{k'}\vect{q}}^{\nu\nu'\omega} = -\beta
    G_{\vphantom{\nu'}\vect{k}\nu} G^{}_{\vphantom{\nu'}(\vect{k}+\vect{q})(\nu+\omega)}
    \delta_{\nu\nu'} \delta_{\vect{k}\vect{k'}}$ the lattice Bethe--Salpeter equation now
    allows us to calculate the generalized DMFT lattice susceptibility
    \begin{equation}
      \label{equ:BSE}
      \chi_{r,\vect{k}\vect{k'}\vect{q}}^{\vphantom{\nu_1}\nu\nu'\omega}=
        \chi_{0,\vect{k}\vect{k'}\vect{q}}^{\vphantom{\nu_1}\nu\nu'\omega}-
        \sum_{\substack{\nu_1\nu_2\\\vect{k_1}\vect{k_2}}}
        \chi_{\vphantom{'}0,\vect{k}\vect{k_1}\vect{q}}^{\nu\nu_1\omega}
        \Gamma_{\vphantom{'}r}^{\nu_1\nu_2\omega}
        \chi_{r,\vect{k_2}\vect{k'}\vect{q}}^{\nu_2\nu'\omega}.
    \end{equation}
    Here and in the following, we distinguish lattice and local quantities through the
    additional momentum indices $\vect{k}$, $\vect{k'}$, $\vect{q}$; and implicitly assume
    a factor $1/\beta$ for each Matsubara frequency sum, as in
    Ref.~\onlinecite{RMPVertex}. That is, $\sum\limits_{\nu_1}$ actually means
    $\frac{1}{\beta} \sum\limits_{\nu_1}$. From these, physical susceptibilities at
    frequency $\omega$ and momentum $\vect{q}$ can be calculated through summing over
    $\nu$, $\nu'$, $\vect{k}$, $\vect{k'}$.

  \subsection{Abinitio\bfdga\ self-energy}
  \label{ssec:ab-initio-dga}
    Similar to \cref{equ:BSE}, we can also calculate the full vertex $F$ through the
    Bethe--Salpeter equation
    \begin{equation}
      \label{equ:F}
      F_{r,\vect{k}\vect{k'}\vect{q}}^{\vphantom{\nu_1}\nu\nu'\omega}=
      \Gamma_{\vphantom{\nu'}r}^{\vphantom{\nu_1}\nu\nu'\omega}+
      \sum_{\substack{\nu_1\nu_2\\\vect{k_1}\vect{k_2}}}
      \Gamma_{\vphantom{\nu'}r}^{\nu\nu_1\omega}
      \chi_{\vphantom{\nu'}0,\vect{k_1}\vect{k_2}\vect{q}}^{\nu_1\nu_2\omega}
      F_{r,\vect{k_2}\vect{k'}\vect{q}}^{\nu_2\nu'\omega}.
    \end{equation}
    However in the ladder approximation the resulting $F$ from \Cref{equ:F} simplifies
    further and does not have an explicit dependence on $\vect{k}$ and $\vect{k'}$:
    \begin{equation}
      \label{equ:Fksum}
      F_{\vphantom{0}r,\vect{q}}^{\vphantom{\nu_1}\nu\nu'\omega}=
      \Gamma_{\vphantom{0}r}^{\vphantom{\nu_1}\nu\nu'\omega}+\sum_{\substack{\nu_1\nu_2}}
      \Gamma_{\vphantom{0}r}^{\nu\nu_1\omega}\chi_{0,\vect{q}}^{\nu_1\nu_2\omega}
      F_{\vphantom{0}r,\vect{q}}^{\nu_2\nu'\omega}.
    \end{equation}
    Here we defined $\chi_{\vphantom{'}0,\vect{q}}^{\nu\nu'\omega} =
    \sum_{\vect{k},\vect{k'}} \chi_{0,\vect{k}\vect{k'}\vect{q}}^{\nu\nu'\omega}$. Please
    note that the generated non-local full vertices $F$ in \Cref{equ:Fksum} are no longer
    crossing symmetric. By taking into account the corresponding diagrams in the
    transversal particle--hole channel we get for the density component
    \begin{eqnarray}
      \label{equ:Fcross}
      F_{d,\vect{k}\vect{k'}\vect{q}}^{\vphantom{()}\nu\nu'\omega} =
      F_{\vphantom{'}d,\vect{q}}^{\vphantom{()}\nu\nu'\omega} + \nonumber \frac{1}{2}
      F_{d,\vect{k'}-\vect{k}}^{\text{nl}\;(\nu'-\nu)(\nu'-\omega)\nu'} \\
      +\frac{3}{2} F_{m,\vect{k'}-\vect{k}}^{\text{nl}\;(\nu'-\nu)(\nu'-\omega)\nu'}
    \end{eqnarray}
    where we defined $F_{r,\vect{q}}^{\text{nl}\;\nu\nu'\omega} =
    F_{r,\vect{q}}^{\nu\nu'\omega} - F_{r}^{\nu\nu'\omega}$. From this vertex we can
    calculate the \dga\ self-energy
    \begin{align}
      \label{equ:EOM}
      \Sigma_{\vphantom{'}\vect{k}\nu}=\frac{Un}{2}-{U}\sum_{\substack{\nu'\omega\\
      \vect{k'}\vect{q}}}&
      F_{d,\vect{kk'q}}^{\nu\nu'\omega}G_{\vect{k'}\nu'}
      G_{(\vect{k'}+\vect{q})(\nu'+\omega)}\nonumber\\
      &\times G_{\vphantom{'}(\vect{k}+\vect{q})(\nu+\omega)},
    \end{align}
    where $n$ is the electron density entering in the Hartree term. The actual
    calculations for both the DMFT susceptibility and the \dga\ self-energy are done using
    the ADGA program package which together with further computational details is
    discussed in Refs.~\onlinecite{Galler2016,adga-cpc}.

  \subsection{Jackknife}
  \label{ssec:jackknife}
    The jackknife is a resampling method, used for bias reduction and error estimation. It
    is a versatile method for a range of problems, hence the name
    \cite{miller-trustworthy-jackknife,miller-jackknife-review}.

    Before we describe the jackknife in detail, let us recapitulate some statistical terms
    that we will need later on. Let $\theta$ be the true, yet unknown, value of some
    quantity. To access $\theta$ we construct an estimator, $\estim\theta$, which is a
    function of $n$ random variables, $X_1, \ldots, X_n$. In our case $X_i$ are just Monte
    Carlo measurements. The bias of the estimator $\estim{\theta}$ is then given by the
    difference between its expectation value $\expect[\estim{\theta}]$ and the true value
    $\theta$,
    \begin{equation}
      \bias[\estim{\theta}]\Def\expect[\estim{\theta}]-\theta.
      \label{equ:bias-definition}
    \end{equation}
    For $\estim\theta$ to be useful, its bias must be systematically improvable by
    increasing the sample size $n$. More specifically, a sequence of estimators
    $\{\estim{\theta}_n(X_1,\dots,X_n)\}$ is called {\em consistent} with $\theta$ if and
    only if the probability of deviating from $\theta$ goes to zero with $n$, i.e.,   
    \begin{equation}
      \lim_{n\rightarrow\infty}\Pr(|\estim{\theta}_n(X_1,\dots,X_n)-\theta|>\epsilon)=0,
      \quad\forall\epsilon>0.
      \label{equ:consistency-definition}
    \end{equation}

    Now let us explain jackknife resampling based on the following general problem. Assume
    we want to calculate some quantity $y=f(x)$, where $f$ is some arbitrary, known
    function. If we knew the true value of $x$ the task would be trivial. In our case,
    however, $x$ is a random variable and we only have access to a sample $\{x_i\}$ of
    size $n$. Therefore we need to find a good estimator $\estim{y}$ and be able to
    quantify its error. In \cref{sec:self-energy,sec:susceptibilities} the input samples
    $x_i$ are the two-particle Green's functions from QMC calculations, $y$ is either the
    \dga\ self-energy or the DMFT susceptibility, $f$ is given by the Bethe--Salpeter
    equation and in case of the self-energy also by the Schwinger--Dyson equation.

    For simplicity, we only propagate the error in the two-particle Green's function. This
    is justified, because the one-particle Green's function is calculated by symmetric
    improved estimators and thus its error is smaller by several orders of
    magnitude\cite{Kaufmann2019}.

    The general strategy of resampling techniques is to generate subsamples from the
    initial one which are preferably independent and identically distributed (iid). With
    this, one can then estimate certain statistics of the sample like its mean or
    variance. In case of the jackknife a bias estimation and reduction is also possible.
    To show this, let $\estim{y}_n$ be a consistent estimator function for $y$. A common
    choice that we used for all calculations in this paper is
    \begin{equation}
      \estim{y}_n(x_1,\dots,x_n)=f(\mean{x}),
      \label{equ:transformed-input-mean}
    \end{equation}
    where $f$ is the function from the original problem and $\mean{x}$ is the input sample
    mean. The latter is an unbiased, consistent estimator for $x$. If $f$ is a continuous
    function, it is easy to see that $f(\mean{x})$ is a consistent estimator for $y$. We
    note that if $f$ is linear, then $f(\mean{x})$ is already unbiased.

    After choosing $\estim{y}_n$ we generate $n$ leave-one-out samples
    \begin{equation}
      y_{-i}=\estim{y}_{n-1}(x_1,\dots,x_{i-1},x_{i+1},\dots,x_n).
      \label{equ:leave-one-out-samples}
    \end{equation}
    Note that this simple choice of resampling implies that the leave-one-out samples are
    also consistent estimators for $y$ and carry a different but similar bias. The reason
    why this is important for obtaining a bias-corrected estimator will become clearer in
    the following. Consistency requires the bias to vanish in the limit of
    $n\rightarrow\infty$. Thus, if $f$ is not only continuous but also analytic around the
    true value $x$, we can expand the bias of $\estim{y}_n$ in powers of $1/n$ and write
    \begin{align}
      \bias[\estim{y}_n]=&\frac{a}{n}+\frac{b}{n^2}+\order(n^{-3}),
      \label{equ:y-hat-bias}\\
      \bias[y_{-i}]=&\bias[\estim{y}_{n-1}]=\frac{a}{n-1}+\frac{b}{(n-1)^2}+
        \order(n^{-3}),
      \label{equ:leave-one-out-bias}
    \end{align}
    with some $a$ and $b$. The expectation value as well as the bias are linear operators
    [see \cref{equ:bias-definition}]. It is therefore easy to construct new samples $y'_i$
    in such a way that we get rid of the leading $\order(1/n)$ term in the bias
    \begin{align}
      y'_i=&n\estim{y}_n-(n-1)y_{-i},
      \label{equ:new-samples}\\
      \bias[y'_i]=&n\bias[\estim{y}_n]-(n-1)\bias[y_{-i}]\nonumber\\
        =&(a+\frac{b}{n})-(a+\frac{b}{n-1})+\order(n^{-3})\nonumber\\
        =&-\frac{b}{n(n-1)}+\order(n^{-3})=\order(n^{-2}).
      \label{equ:new-sample-bias}
    \end{align}
    By calculating the sample mean of the $y'_i$ we finally arrive at the bias-corrected
    jackknife estimator
    \begin{equation}
      \estim{y}_\text{JK}=\mean{y}'=\frac{1}{n}\sum_{i=1}^n y'_i.
      \label{equ:y-jk}\\
    \end{equation}
    Its expectation value is given by
    \begin{equation}
      \begin{split}
        \expect[\estim{y}_\text{JK}]=&\frac{1}{n}\sum_{i=1}^n\Big(\expect[y'_i]\Big)
        =\frac{1}{n}\sum_{i=1}^n\Big(n\expect[\estim{y}_n]-(n-1)\expect[y_{-i}]\Big)\\
        =&\frac{1}{n}\sum_{i=1}^n\Big(n(y+\bias[\estim{y}_n])-
        (n-1)(y+\bias[y_{-i}])\Big)\\
        =&\frac{1}{n}\sum_{i=1}^n\Big(y+\bias[y'_i]\Big)=y+\order(n^{-2}),
        \label{equ:y-jk-expectation}
      \end{split}
    \end{equation}
    where \cref{equ:new-sample-bias} was used in the last step. This shows that
    $\estim{y}_\text{JK}$ is a consistent estimator for $y$ with a reduced bias of
    $\order(1/n^2)$.

    Without proof, consistent estimators for the variance, standard deviation, covariance,
    etc.~of $y$ can also be obtained by calculating the corresponding sample estimates of
    $\{y'_i\}$\cite{efron-jackknife}. In
    \cref{sec:2-particle-greens-function,sec:self-energy,sec:susceptibilities} we are
    specifically interested in the standard error of the mean ($\text{SEM}$) and the
    linear correlation, $\corr[\cdot,\cdot]$. The latter is estimated by the sample
    Pearson correlation coefficients $r$. For scalar random variables $p$ and $q$ with
    samples $\{p_i\}$ and $\{q_i\}$ of size $n$ the following holds
    \begin{align}
      \text{SEM}_p\Def&\;\frac{s_p}{\sqrt{n}}=
        \sqrt{\frac{1}{n(n-1)}\sum_{i=1}^n\left|p_i-\mean{p}\right|^2}
      \label{equ:sem-definition}\\
      \wideestim{\corr}[p,q] =& \;r_{pq} =
      \frac{\sum_{i=1}^n(p_i-\mean{p})(q_i-\mean{q})^\ast}
      {\sqrt{\sum_{i=1}^n|p_i-\mean{p}|^2}\sqrt{\sum_{i=1}^n|q_i-\mean{q}|^2}}.
      \label{equ:corr-definition}
    \end{align}
    Here $s$ denotes the corrected sample standard deviation, $\mean{p}$ the sample mean
    of the $p_i$, and $\wideestim{\corr}$ the estimated correlation. The generalization to
    random vectors and objects of higher rank is straightforward by componentwise
    application of the above formulas. In the following, $p_i$ is a scalar component of
    the $i$th sample $x_i$ or $y_i$, e.g., $p_i=(\Sigma_{\text{\dga}, \vect{k}=(0,0),
    \nu=\pi/\beta})_i$ might be the $i$th measurement of the self-energy at a fixed
    momentum ${\mathbf k}$ and frequency $\nu$.
    
    Let us note an important caveat in using \cref{equ:corr-definition} as estimator for a
    $k\times k$ correlation matrix $r_{pq}$ with a large number of features $k$: While
    each component of the covariance converges as $1/\sqrt{n}$ regardless of $k$, the
    eigenvalues of the covariance matrix, which are used to construct independent errors,
    converge only as a function of $n/k$. In particular, the estimator yields a singular
    correlation matrix for any $n<k$.

    For practical use, the whole derivation and discussion of the jackknife above can be
    condensed into three simple steps:
    \begin{enumerate}
      \item Resample
        \begin{equation}
          x_i\rightarrow x'_i=\frac{1}{n-1}\sum_{j\ne i}x_j=\frac{n\mean{x}-x_i}{n-1}
          \label{equ:resample}
        \end{equation}
      \item Transform
        \begin{equation}
          y'_i=nf(\mean{x})-(n-1)f(x'_i)
          \label{equ:transform}
        \end{equation}
      \item Calculate sample statistics of $\{y'_i\}$, e.g.,
        \cref{equ:y-jk,equ:sem-definition,equ:corr-definition}
    \end{enumerate}

    Another statistical method, similar to the jackknife, is the bootstrap. It is more
    powerful but usually requires a greater number of resamples to take advantage of
    that\cite{efron1979}. Depending on the specific problem, at least hundreds or
    thousands of new samples are drawn for the bootstrap
    method\cite{efron-jackknife,wilcox2010} as opposed to the 16 to 256 jackknife samples
    that are used in \cref{sec:self-energy,sec:susceptibilities}. Since each resample
    requires a full \dga\ calculation, the jackknife is computationally cheaper and
    therefore the method of choice.

  \subsection{Parallel implementation}
  \label{ssec:parallel-implementation}
    The main focus of this paper is on the jackknife estimates of the self-energy and
    susceptibilities calculated within ADGA. In this case the parallelization is simple,
    because the ADGA calculation is by far the most computationally intensive task and
    already implemented in a parallel way. Therefore the actual jackknife part is
    programmed in serial and only the calls to the ADGA code are done in parallel.
  
\section{Statistical analysis of the input: two-particle Green's function}
\label{sec:2-particle-greens-function}
  Before we analyze the \dga\ self-energy and DMFT susceptibilities, let us take a closer
  look at the input of the DMFT and \dga\ calculations, namely the two-particle Green's
  function $G^{(2),\nu\nu'\omega}$. In particular we want to check if the correlations of
  the self-energy and susceptibilities are completely intrinsic or if they originate at
  least in part from the input. For this reason we estimate
  $\corr[G_{\uparrow\uparrow}^{(2),\nu_1\nu'_1\omega_1},
  G_{\uparrow\uparrow}^{(2),\nu_2\nu'_2\omega_2}]$ for various frequency combinations, and
  plot two-dimensional cuts of this high-dimensional quantity.

  All QMC simulations were done for the 2D square lattice Hubbard model at half-filling
  using the following parameters: $U=4t$, $\beta=\{2/t,4/t\}$, where the hopping amplitude
  $t=1$ serves as our energy unit. The hopping matrix in \Cref{equ:lat} is taken to permit
  only nearest-neighbor hopping. The number of fermionic frequencies is 40 for $\beta=2$
  and 80 for $\beta=4$. Due to the imposed particle-hole symmetry, the two-particle
  Green's function and therefore also its correlation matrix is purely real. Before the
  latter was estimated, the total number of $n_t$ QMC measurements were divided equally
  into $n_b$ bins. The $n_m=n_t/n_b$ measurements in each bin were then averaged and used
  as the samples for the estimations. All results in this section were obtained with
  $n_m=2.4\times 10^6$.

\begin{figure}
    \includegraphics[width=86mm]{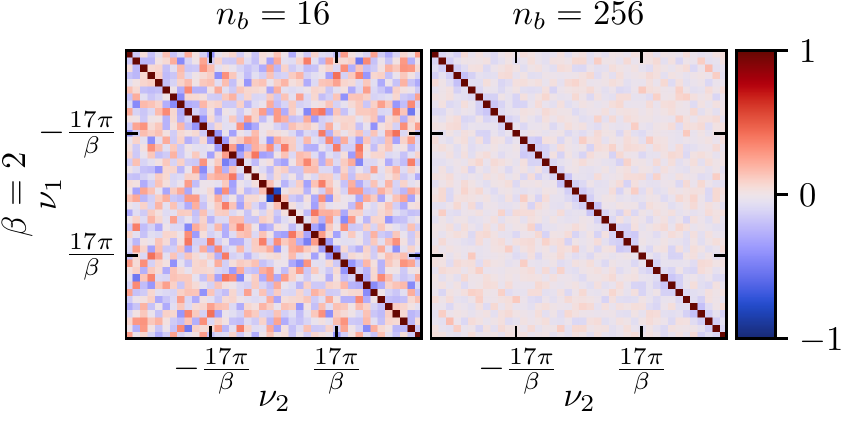}
    \includegraphics[width=86mm]{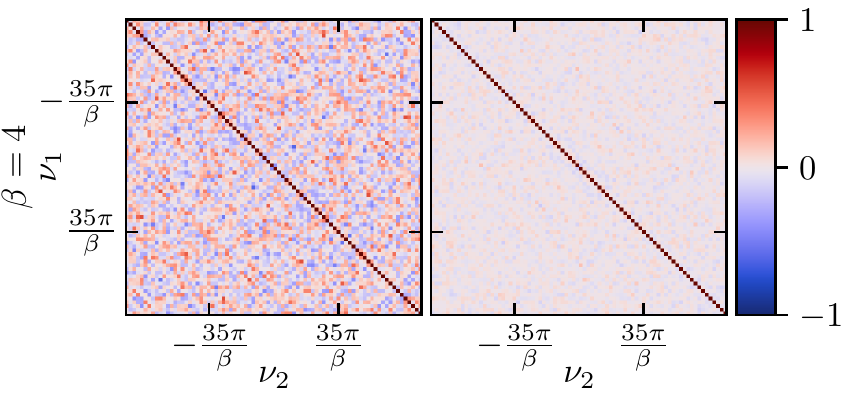}
    \caption{Estimated correlation matrix
    $\wideestim{\corr}[G^{(2)}_\text{cut1}(\nu_1),G^{(2)}_\text{cut1}(\nu_2)]$ for the
    two-particle Green's function at the cut $\nu=\nu'=\nu_i$, $\omega=0$ comparing
    different temperatures and numbers of bins. The $40\times 40$ matrices in the top row
    and the $80\times 80$ matrices in the bottom row correspond to 40 and 80 fermionic
    frequencies, respectively.}
    \label{fig:g4-nu-nu-correlations}
  \end{figure}

  In \cref{fig:g4-nu-nu-correlations} the estimated correlation of
  \begin{equation}
    G^{(2)}_\text{cut1}(\nu_1)\Def
    G^{(2),\nu=\nu_1,\nu'=\nu_1,\omega=0}_{\uparrow\uparrow}
    \label{equ:g-cut-1}
  \end{equation}
  with itself is shown for two temperatures $\beta=2$ (top) and $\beta=4$ (bottom) and two
  numbers of bins $n_b=16$ (left) and $n_b=256$ (right).
  Although there is quite some noise in the plots with fewer bins, the diagonal matrix
  structure is evident in all four graphs (note the sign of the $y$-axis resembling the
  typical arrangement in a matrix). This suggests that the two-particle Green's function
  is uncorrelated at different frequencies, at least along the cut.
  
\begin{figure}
    \centering
    \includegraphics[width=86mm]{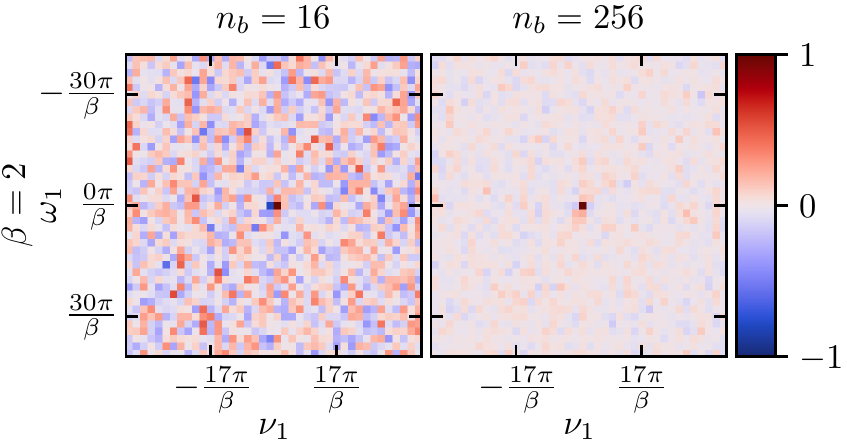}
    \caption{Same as \cref{fig:g4-nu-nu-correlations} (top) but now for the estimated
    correlation matrix $\wideestim{\corr}[G^{(2)}_\text{cut2}(\omega_1),
    G^{(2)}_\text{cut1}(\nu_1)]$, i.e., the correlation between cut2 with $\omega_1$,
    $\nu'=\nu=0$ and cut1 with $\omega=0$, $\nu'=\nu=\nu_1$. The $41\times40$ correlation
    matrices correspond to 41 bosonic and 40 fermionic frequencies. As in
    \cref{fig:g4-nu-nu-correlations}, for sufficiently many bins (right) the off-diagonal
    components of the correlation matrix vanish. (Note that in this figure there are no
    diagonal components, since there are different frequencies on the axes.)}
    \label{fig:g4-omega-nu-correlations-b2}
  \end{figure}

\begin{figure}
    \centering
    \includegraphics[width=86mm]{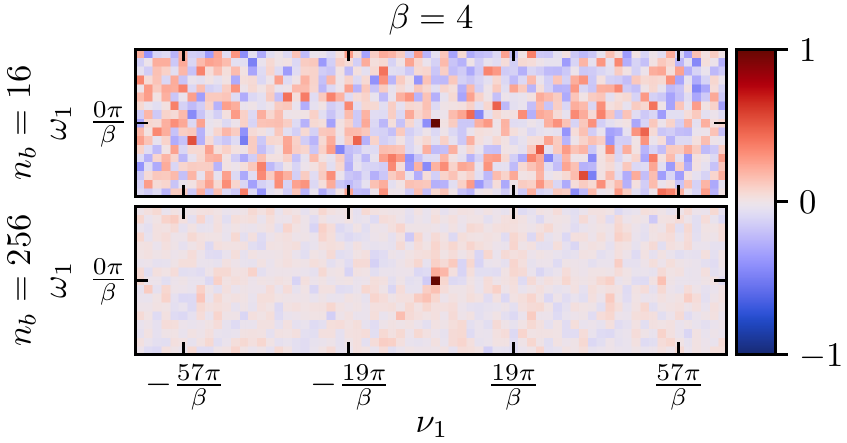}
    \caption{Same as \cref{fig:g4-omega-nu-correlations-b2} but now at $\beta=4$. The
    $17\times 70$ matrices correspond to 70 fermionic and 17 bosonic frequencies (which we
    reduced to save computational time as there was no measurable change in the
    self-energy).}
    \label{fig:g4-omega-nu-correlations-b4}
  \end{figure}

  To further test this, we next consider the estimated correlation between
  \begin{equation}
    G^{(2)}_\text{cut2}(\omega_1)\Def
    G^{(2),\nu=0,\nu'=0,\omega=\omega_1}_{\uparrow\uparrow}
    \label{g-cut-2}
  \end{equation}
  and $G^{(2)}_\text{cut1}$ which is shown for $\beta=2$ in
  \cref{fig:g4-omega-nu-correlations-b2} and for $\beta=4$ in
  \cref{fig:g4-omega-nu-correlations-b4}.
  The same numbers of bins are used as before. Apart from the noise, the correlation
  vanishes again.

  Altogether, this suggests that it is reasonable to assume that the two-particle Green's
  function at different frequencies as calculated with QMC is uncorrelated. Since this is
  the input to the subsequent DMFT or \dga\ calculations, any correlations of the output
  data must emerge through the post-processing.

\section{Self-energy}
\label{sec:self-energy}
  From the local two-particle Green's function as well as the local one-particle Green's
  function as an input, the self-energy $\Sigma_{{\text\dga},\nu\vect{k}}$ of the 2D
  square lattice Hubbard model at half-filling is calculated in \dga. All results in this
  section are generated using the same parameters as in
  \cref{sec:2-particle-greens-function}: $U=4t$, $\beta=\{2/t,4/t\}$, $t=1$, $t'=0$, i.e.,
  only nearest neighbor hopping is considered. The \dga\ calculations use nine bosonic as
  well as 40 and 80 fermionic frequencies for $\beta=2$ and $\beta=4$, respectively. The
  inner momentum-($\vect{k}$-)grid is $48\times48$ and the transfer
  momentum-($\vect{q}$-)grid is $12\times12$ for $\beta=2$ and $24\times24$ for $\beta=4$.
  Before jackknife resampling is applied, the total number $n_t$ of QMC measurements of
  the two-particle Green's function is divided equally into $n_b$ bins with $n_m=n_t/n_b$
  measurements per bin which are averaged for each bin. These $n_b$ averages are then used
  as the input samples for the jackknife. For a more compact notation and easier
  comparison of the multi panel figures in this section it is convenient to give the
  number of measurements as a multiple of $n_0=2.4\times 10^6$.

  \subsection{Standard error of the mean (SEM)}
  \label{ssec:self-energy-sem}
\begin{figure}
      \centering
      \includegraphics[width=86mm]{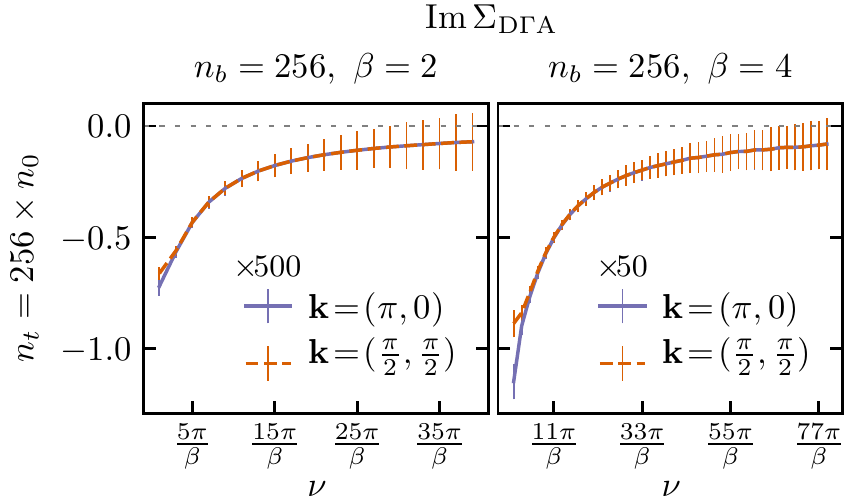}
      \caption{Imaginary part and jackknife SEM of the \dga\ self-energy at different
      temperatures. For better visibility, the errors bars are enlarged by a factor of 500
      on the left and by 50 on the right.}
      \label{fig:2-self-energies}
    \end{figure}

    \Cref{fig:2-self-energies} shows the imaginary part of the \dga\ self-energy at
    $\beta=2$ and $\beta=4$ using $256$ bins with $n_0$ QMC measurements each.
    Note that the error bars are enlarged by a factor of 500 for the left and 50 for the
    right plot. Taking this rescaling of the error into account, the SEM for $\beta=4$
    (right) is about 10 times higher than for $\beta=2$ (left) but still very small. We
    do not plot the real part of the self-energy because on the Fermi surface at
    half-filling it is just a constant.
    
\begin{figure}
      \centering
      \includegraphics[width=86mm]{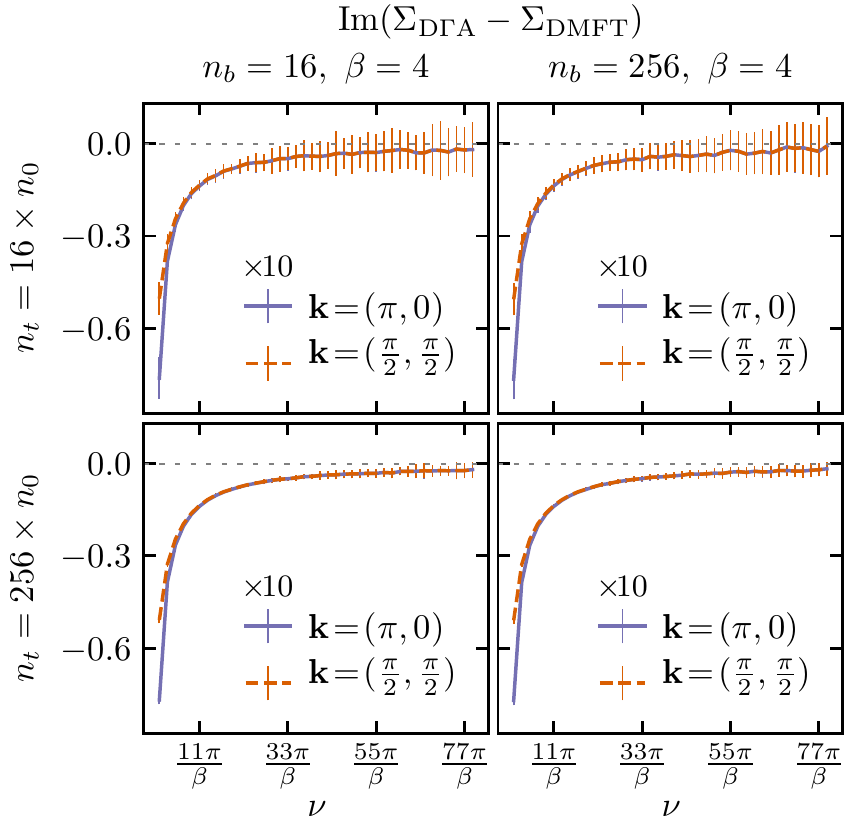}
      \caption{Imaginary part and jackknife SEM of the difference between the DMFT and
      \dga\ self-energy, $\Sigma_{\text{\dga},\vect{k}\nu}-\Sigma_{\text{DMFT},\nu}$, at
      $\beta=4$. Different numbers of bins and total measurements are compared. For better
      visibility the error bars are enlarged by a factor of 10.}
      \label{fig:2p-correlations-b4}
    \end{figure}

    Since ADGA actually calculates two-particle corrections to the DMFT self-energy it is
    more reasonable to plot the error bars of the difference between the DMFT and \dga\
    self-energy, $\Sigma_{\text{\dga},\vect{k}\nu}-\Sigma_{\text{DMFT},\nu}$. In
    \cref{fig:2p-correlations-b4} the imaginary part of this \dga\ self-energy correction
    is plotted for $\beta=4$ and various combinations of the total number of QMC
    measurements $n_t$ and number of bins $n_b$.
    In the top (bottom) row $n_t=16\times n_0$ ($256\times n_0$); in the left (right)
    column $n_b=16$ (256) bins. It is obvious that the SEM scales with $n_t$. That is, the
    error for the larger number of measurements $n_t$ (bottom row in
    \cref{fig:2p-correlations-b4}) is smaller than for a smaller $n_t$ (top row in
    \cref{fig:2p-correlations-b4}), just as expected. However, there is practically no
    dependence of the error on the number of bins $n_b$ (left vs~right column of
    \cref{fig:2p-correlations-b4}). One can also see that even only $16\times n_0\approx
    4\times 10^7$ total measurements (top row in \cref{fig:2p-correlations-b4}) lead to
    still acceptable sizes of the error bars (note they are resized by a factor of ten).

  \subsection{Correlation matrix}
  \label{ssec:self-energy-correlations}
\begin{figure}
      \centering
      \includegraphics[width=86mm]{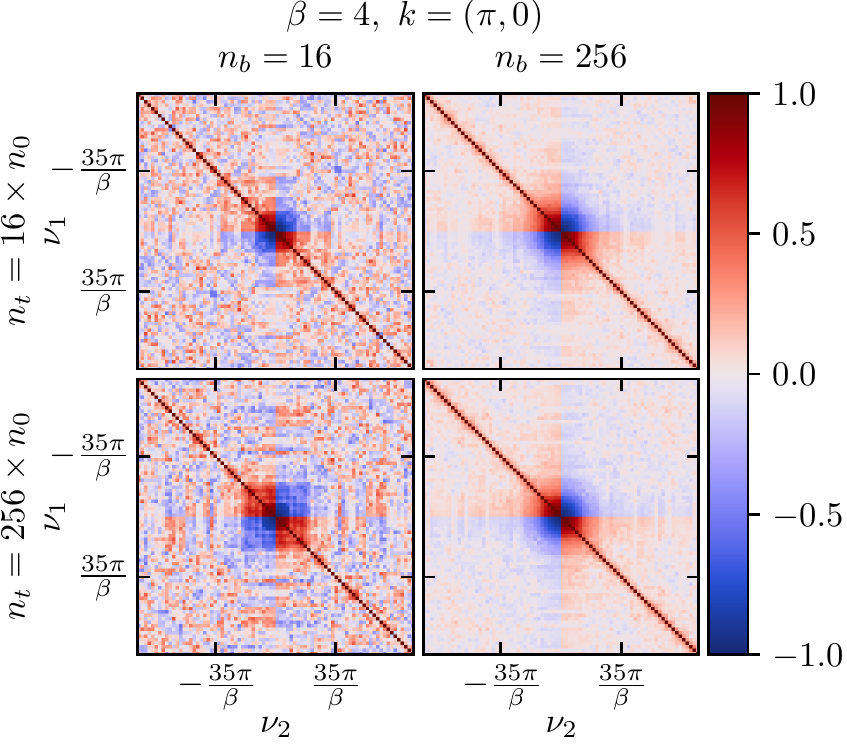}
      \caption{Real part of the estimated correlation matrix
      $\wideestim{\corr}[\Sigma_{{\text\dga},\vect{k}\nu_1},
      \Sigma_{{\text\dga},\vect{k}\nu_2}]$. As in \cref{fig:2p-correlations-b4} different
      numbers of bins and total measurements are compared. The $80\times80$ matrices
      correspond to 80 fermionic frequencies.}
      \label{fig:siw-nu-nu-correlations-b4}
    \end{figure}

    Let us now turn to the correlation of the different frequency components of the \dga\
    self-energy. \Cref{fig:siw-nu-nu-correlations-b4} shows the real part of the estimate
    for the correlation of $\Sigma_{{\text\dga},\vect{k} \nu}$ with
    $\Sigma_{{\text\dga},\vect{k} \nu'}$ for $\beta=4$ and $\vect{k}=(\pi,0)$. (For the
    statistical analysis of the input data, we refer the reader to
    \cref{sec:2-particle-greens-function}.)
    It uses the same layout as \cref{fig:2p-correlations-b4}, i.e., $n_t=16\times n_0$ in
    the top row and $n_t=256\times n_0$ in the bottom one, with $n_b=16$ on the left and
    $n_b=256$ on the right. Contrary to the SEM, the estimated correlation matrix strongly
    depends on $n_b$ as the comparison between the left and right column in
    \cref{fig:siw-nu-nu-correlations-b4} shows. While increasing $n_t$ does improve the
    noise slightly, a large number of jackknife samples is crucial for an acceptable noise
    level. It is evident in all four plots that the largest correlations appear in the
    low-frequency region. Disregarding noise, the only correlations outside of this area
    are those between low and high frequencies.

\begin{figure}
      \centering
      \includegraphics[width=86mm]{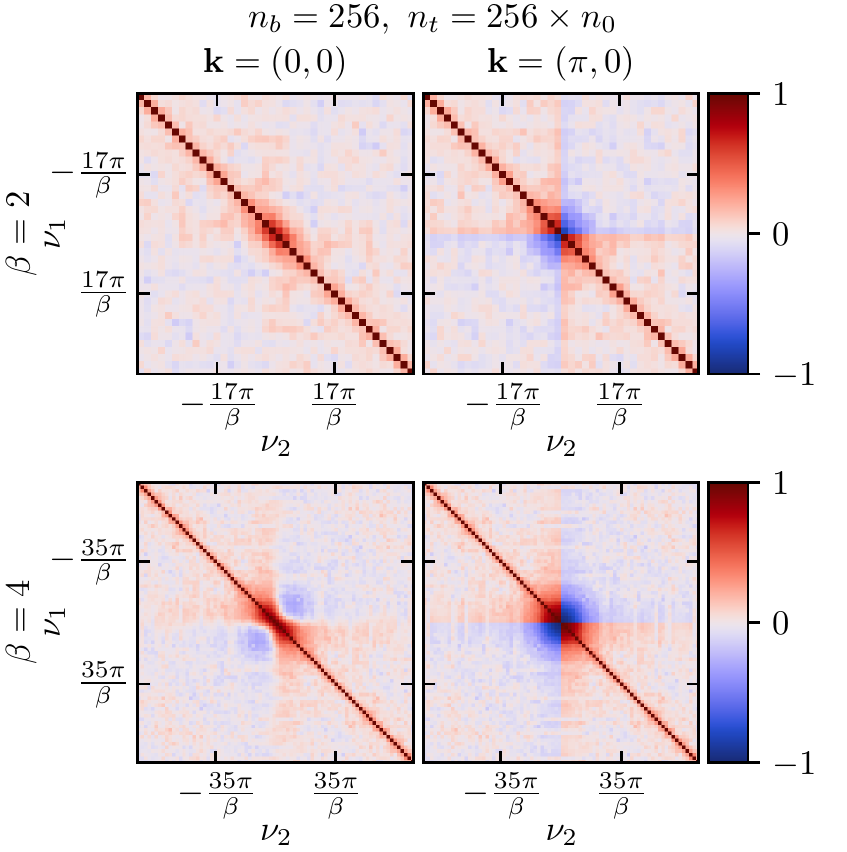}
      \caption{Real part of the estimated correlation matrix
      $\wideestim{\corr}[\Sigma_{\text{\dga},\vect{k}\nu_1},
      \Sigma_{\text{\dga},\vect{k}\nu_2}]$ at different temperatures and
      $\vect{k}$-points. The $40\times40$ (top) and $80\times80$ matrices (bottom)
      correspond to 40 and 80 fermionic frequencies, respectively.}
      \label{fig:siw-nu-nu-correlations-b2-b4}
    \end{figure}

    \Cref{fig:siw-nu-nu-correlations-b2-b4} shows the dependence of the real part of the
    estimated correlation matrix on $\beta$ (from the top to the bottom of
    \cref{fig:siw-nu-nu-correlations-b2-b4}) and $\vect{k}$ (from left to right).
    One can see that the correlation increases slightly with increasing $\beta$ (or
    decreasing temperature) and is also more pronounced for $\vect{k}=(\pi,0)$ (right)
    than for $\vect{k}=(0,0)$ (left). However, the general structure of the matrix -- high
    correlation at low frequencies, very low correlation otherwise -- is similar in all
    cases.

    We do not plot the imaginary part of the estimated correlation matrix because with the
    chosen parameters (half-filling and $\vect{k}$-points on Fermi surface) it vanishes
    and shows only noise.

    The cross-correlation between frequencies of the D$\Gamma$A self-energy can be
    understood from \Cref{equ:EOM}. It can be rewritten, such that we have the DMFT
    self-energy plus non-local corrections arising from the two-particle Green's
    function\cite{Galler2016}. Both DMFT and D$\Gamma$A self-energy have the same
    asymptotic behavior, thus non-local corrections have to be restricted to the lower
    Matsubara frequencies. Since here we consider the correlation arising from the
    two-particle Green's function, we can conclude that they mainly influence the
    low-frequency region. Furthermore, in the model under consideration vertex corrections
    increase with lower temperatures, and the largest influence is to be expected at
    $\vect{k}=(\pi,0)$, where the formation of a pseudo-gap can be observed.

    The symmetries in
    \cref{fig:siw-nu-nu-correlations-b4,fig:siw-nu-nu-correlations-b2-b4} can be explained
    by the definition of the estimated correlation coefficients and properties of the
    self-energy. Using \cref{equ:corr-definition} and $\Sigma(\nu)^\ast=\Sigma(-\nu)$
    yields
    \begin{equation}
    \begin{split}
      \wideestim{\corr}[\Sigma(\nu_1),\Sigma(\nu_2)]
      =&\;\wideestim{\corr}[\Sigma(\nu_2),\Sigma(\nu_1)]^\ast\\
      =&\;\wideestim{\corr}[\Sigma(\nu_2)^\ast,\Sigma(\nu_1)^\ast]\\
      =&\;\wideestim{\corr}[\Sigma(-\nu_2),\Sigma(-\nu_1)].
      \label{equ:corr-symmetries}
    \end{split}
    \end{equation}
    Therefore the real part of the correlation matrices is symmetric around the main- and
    antidiagonal.

\section{Susceptibilities}
\label{sec:susceptibilities}
  For periodic systems, the density and magnetic DMFT susceptibilities
  $\chi_d(\i\omega,\vect{q})$ and $\chi_m(\i\omega,\vect{q})$ are the Fourier transform of
  the following imaginary time expectation values:
  \begin{align}
    \chi_{\vphantom{'}d}(\tau,l-l')&=\braket{T_{\vphantom{'}\tau}
    (n_{\vphantom{'}l\uparrow}+n_{\vphantom{'}l\downarrow})(\tau)
    (n_{l'\uparrow}+n_{l'\downarrow})(0)},
    \label{equ:chi-d-in-tau}\\
    \chi_{\vphantom{'}m}(\tau,l-l')&=\braket{T_{\vphantom{'}\tau}
    (n_{\vphantom{'}l\uparrow}-n_{\vphantom{'}l\downarrow})(\tau)
    (n_{l'\uparrow}-n_{l'\downarrow})(0)}.
    \label{equ:chi-m-in-tau}
  \end{align}
  Here $l$ and $l'$ are lattice site indices, $T_\tau$ is the time-ordering operator and
  $n_{l\sigma}$ is the electron density at site $l$ with spin
  $\sigma\in\{\uparrow,\downarrow\}$. In DMFT, they are calculated in momentum space for
  the square-lattice Hubbard model at half-filling, using the Bethe--Salpeter equations
  discussed in \cref{ssec:DMFTimp}. All results in this section are generated using the
  same parameters as in \cref{sec:2-particle-greens-function}: $U=4t$, $\beta=\{2/t,
  4/t\}$, where $t=1$ sets the energy unit, and only nearest-neighbor hopping is
  considered. Again, we use nine bosonic as well as 40 and 80 fermionic frequencies for
  $\beta=2$ and $\beta=4$, respectively. The inner momentum- or $\vect{k}$-grid (for the
  one-particle quantities and $\chi_0$) is $48\times48$, whereas the transfer momentum- or
  $\vect{q}$-grid is $12\times12$ for $\beta=2$ and $24\times24$ for $\beta=4$. As before,
  the total number of QMC measurements $n_t$ for the two-particle Green's function is
  given in multiples of $n_0=2.4\times 10^6$ and divided into $n_b$ bins, with
  $n_m=n_t/n_b$ measurements averaged per bin. These averages are then used as the input
  samples for the jackknife. Note that at half-filling, the susceptibilities and therefore
  also their correlation matrices are purely real.

  \subsection{Standard error of the mean (SEM)}
  \label{ssec:susceptibility-error}
\begin{figure}
      \includegraphics[width=86mm]{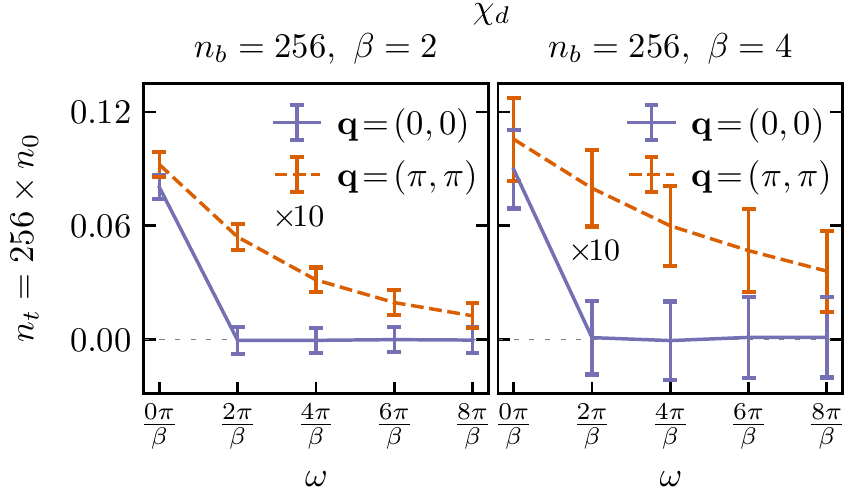}
      \\\vspace{2mm}
      \includegraphics[width=86mm]{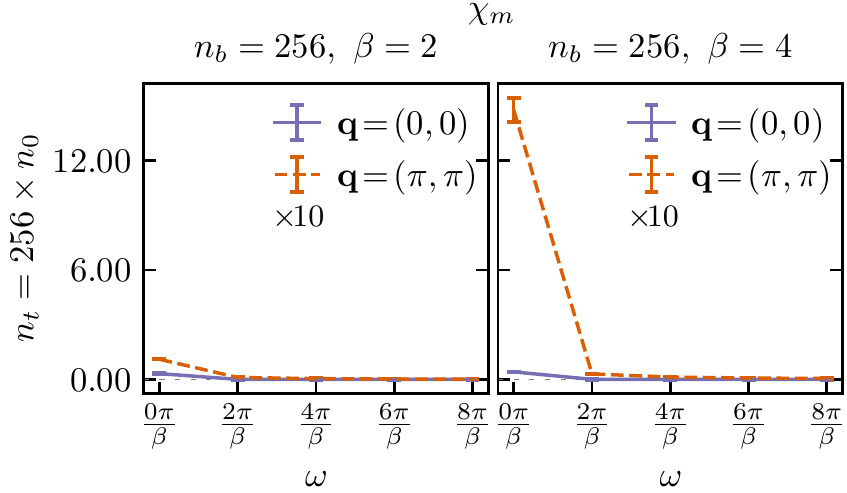}
      \caption{Density and magnetic DMFT susceptibility, $\chi_d$ and $\chi_m$, at
      different temperatures and momenta. For better visibility the error bars are
      enlarged by a factor of 10.}
      \label{fig:susceptibilities-d-m-b2-b4}
    \end{figure}

    \Cref{fig:susceptibilities-d-m-b2-b4} shows the density and magnetic susceptibilities
    at $\beta=2$ and $\beta=4$, where $n_t=256\times n_0$ measurements are
    divided into $n_b=256$ bins.
    Since there is a transition to an antiferromagnetic phase at
    $\beta=4.3$\cite{Kunes2011}, $\chi_m(\vect{q}=(\pi,\pi))$ as well as its error are
    dominated by the contributions at $\omega=0$. This makes it harder to compare the
    plots of the two susceptibilities but one can still see that the SEM increases with
    $\beta$ (left vs~right panels) for both quantities, just like in the case of the
    \dga\ self-energy.

\begin{figure}
      \includegraphics[width=86mm]{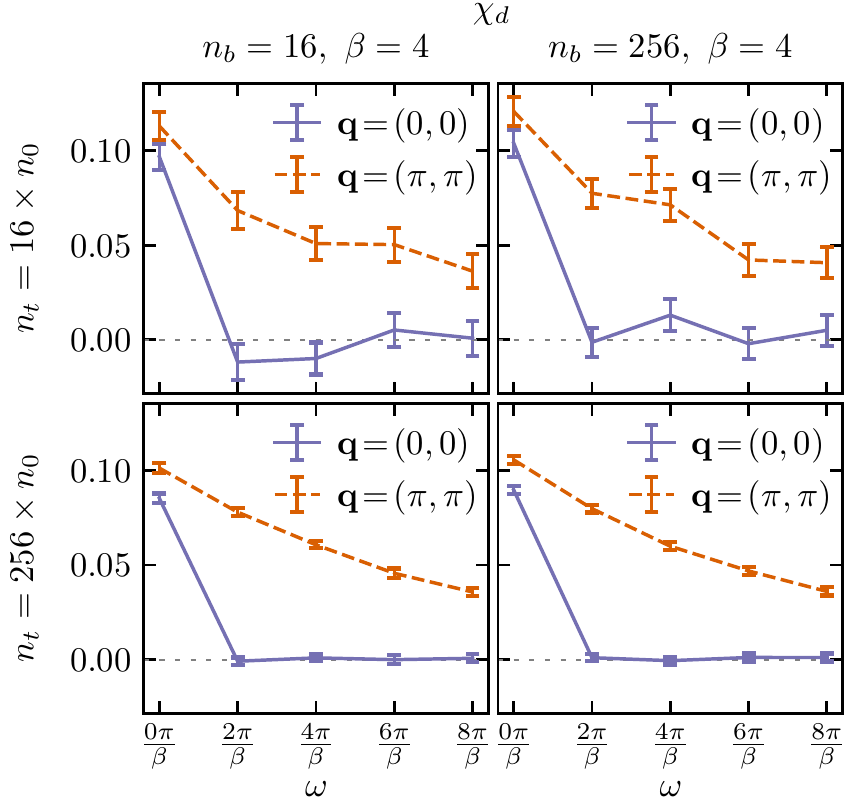}
      \caption{DMFT density susceptibility $\chi_{d}$ at $\beta=4$ for two different
      momenta $\vect{q}$, comparing different numbers of bins and total measurements.}
      \label{fig:susceptibilities-d-b4}
    \end{figure}

\begin{figure}
      \includegraphics[width=86mm]{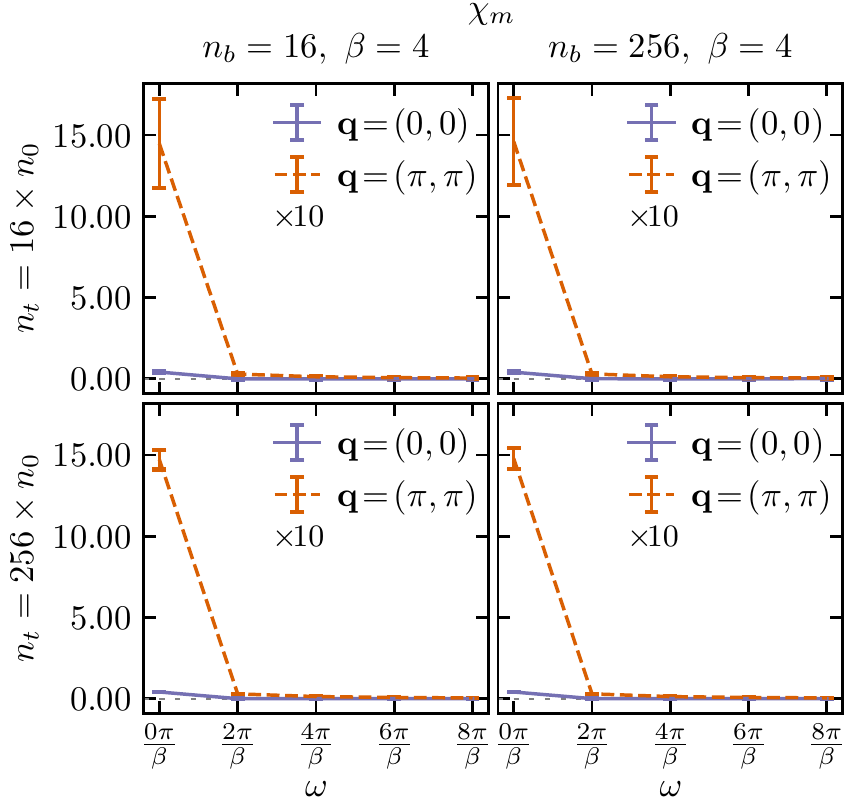}
      \caption{Same as \cref{fig:susceptibilities-d-b4} but for the magnetic
      susceptibility $\chi_m$. For better visibility the error bars are enlarged by a
      factor of 10.}
      \label{fig:susceptibilities-m-b4}
    \end{figure}

    The dependence on the total number of measurements $n_t$ and the number of bins $n_b$
    is shown in \cref{fig:susceptibilities-d-b4} for $\chi_d$ and in
    \cref{fig:susceptibilities-m-b4} for $\chi_m$; both at $\beta=4$.
    They use the same layout as \cref{fig:2p-correlations-b4}, i.e., $n_t=16\times n_0$ in
    the top rows and $256\times n_0$ in the bottom ones, with $n_b=16$ on the left and 256
    on the right. Only the error bars of $\chi_m$ are enlarged by a factor of 10. Similar
    to \cref{sec:self-energy} the SEM scales with the total number of measurements (top
    vs.~bottom panels), but it does not depend on the number of jackknife samples $n_b$
    (left vs.~right panels). Contrary to the results of the self-energy, even using a
    total of $256\times n_0\approx 6\times 10^8$ measurements only yields borderline
    acceptable error bars. This means the main features of the $\chi_d$ curve are still
    recognizable but larger error bars would render the signal statistically
    insignificant. Therefore one should aim for at least $\order(10^9)$ total measurements
    in this case.

  \subsection{Correlation matrix}
  \label{ssec:susceptibility-correlations}
\begin{figure}
      \centering
      \includegraphics[width=86mm]{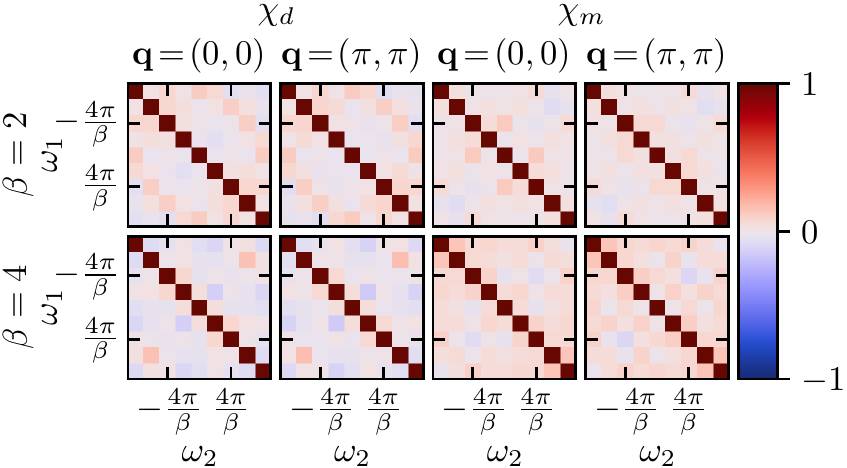}
      \caption{Estimated correlation matrix of the DMFT susceptibilities
      $\wideestim{\corr}[\chi_r(\omega_1),\chi_r(\omega_2)]$, $r\in\{m,d\}$ at different
      temperatures and momenta. A total number of $256\times n_0$ measurements are binned
      into 256 jackknife samples. The $9\times9$ matrices correspond to 9 bosonic
      frequencies.}
      \label{fig:chi-omega-omega-correlations}
    \end{figure}

\begin{figure}
      \centering
      \includegraphics[width=86mm]{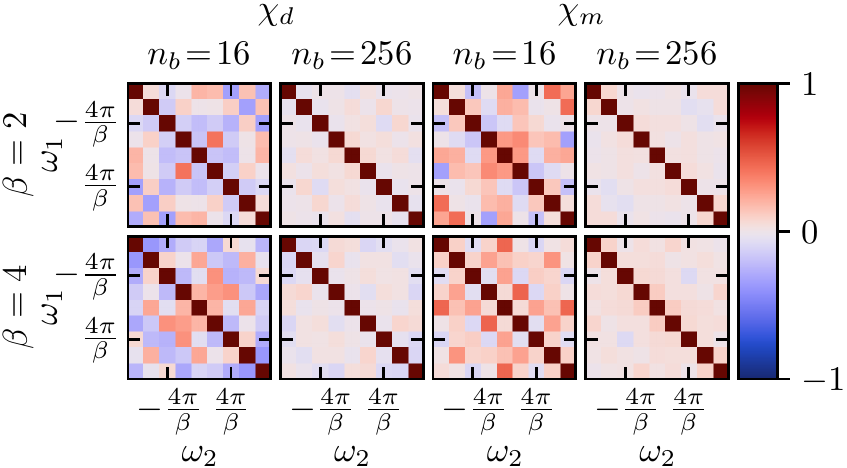}
      \caption{Same as \cref{fig:chi-omega-omega-correlations} but now at
      $\vect{q}=(\pi,0)$ and comparing different numbers of jackknife bins $n_b$ for a
      total number of $16\times n_0$ measurements.}
      \label{fig:chi-omega-omega-correlations-q0006}
    \end{figure}

    The estimate for the correlation of the susceptibilities with themselves is shown in
    \cref{fig:chi-omega-omega-correlations}.
    Both $\chi_d$ and $\chi_m$ are uncorrelated for both temperatures ($\beta=2$ and
    $\beta=4$) and for both momenta [$\vect{q}=(0,0)$ and $\vect{q}=(\pi,\pi)$]. The same
    is true for $\vect{q}=(\pi,0)$, shown in
    \cref{fig:chi-omega-omega-correlations-q0006}, which studies the effects of using more
    jackknife bins.
    Clearly, a good estimate of the correlation matrix of the susceptibility requires more
    than $\order(10)$ bins.

    The symmetries of the correlation matrices in
    \cref{fig:chi-omega-omega-correlations,fig:chi-omega-omega-correlations-q0006} are the
    same as those in \cref{ssec:self-energy-correlations}. They are symmetric around the
    main- and antidiagonal because $\chi_{m,d}(\omega)=\chi_{m,d}(-\omega)$, which is the
    same property as that of the self-energy considering that the susceptibilities are
    also real.

\section{Maximum entropy analytic continuation}
\label{sec:analytic-continuation}
  Extracting real-frequency information, $A(\omega)$, out of Matsubara-frequency data,
  $y(\nu)$, amounts to solving the following fitting problem:
  \begin{equation}
      \min_A \left\Vert y(\nu) - \int \dif\omega K(\nu, \omega) A(\omega) \right\Vert
      = \min_{\vect{A}} || \vect{y} - K \vect{A} || ,
      \label{eq:kernel}
  \end{equation}
  where $K$ is an integral kernel which is different for bosonic or fermionic functions.
  $\vect{y}$ and $\vect{A}$ are the Fourier coefficients of $y(\nu)$ and $A(\omega)$ in an
  appropriate basis. \Cref{eq:kernel} is minimal if and only if the log-likelihood:
  \begin{equation}
    \label{equ:fit-deviation}
    L[\vect{A}] = -\tfrac12 \big(\vect{y} - K\vect{A}\big)^\dagger C^{-1}
                    \big(\vect{y} - K\vect{A}\big),
  \end{equation}
  is maximal, where $C$ is the covariance matrix (a positive definite symmetric matrix).
  \Cref{eq:kernel,equ:fit-deviation} are ill-posed on numerical data as the singular
  values of $K$ drop super-exponentially.

  The maximum entropy method (MEM) \cite{Jarrell1996} is a widely employed method to
  regularize this problem. Briefly, instead of \cref{eq:kernel}, we maximize an augmented
  functional:
  \begin{equation}
    \label{equ:loss-functional}
    Q_\alpha[\vect{A}] = L[\vect{A}] + \alpha S[\vect{A}|\vect{A}_0],
  \end{equation}
  where $S[\vect{A}|\vect{A}_0]$ is the relative (information) entropy with respect to an
  \emph{a priori} default model $\vect{A}_0$. This term regularizes the optimization and
  has to be scaled by a hyperparameter $\alpha$. \Cref{equ:loss-functional} can be used on
  numerical data: $\vect{y}$ is now the sample mean and $C$ is the sample covariance
  matrix in \cref{equ:fit-deviation}.

  \Cref{equ:fit-deviation} can be evaluated much more efficiently, if the covariance
  matrix is diagonal. However, if that is not the case, we may still perform the
  transformation $C = \mathcal{U}^\dagger V \mathcal{U}$, where $\mathcal{U}$ is unitary
  and $V$ is a positive definite diagonal matrix. \Cref{equ:fit-deviation} then acquires
  the simpler form
  \begin{equation}
    \label{equ:fit-deviation-diagonal}
    L[\vect{A}] = -\frac 12 \sum_j
      \frac{\big|\tilde{\vect{y}}_j - (\tilde{K}\vect{A})_j\big|^2}{V_j}
  \end{equation}
  with $\tilde{\vect{y}} = \mathcal{U}\vect{y}$ and $\tilde{K} = \mathcal{U}K$.

  If this rotation of the data and the kernel is done as a pre-processing step, then the
  remaining problem is identical to the case where the covariance matrix is diagonal. In
  particular, it is still possible to treat real and imaginary part as separate variables
  by stacking $\tilde{\vect{y}} \rightarrow [\re\tilde{\vect{y}},\im\tilde{\vect{y}}]$ and
  $\tilde{K} \rightarrow [\re\tilde{K},\im\tilde{K}]$.
  \begin{figure}
    \includegraphics[width=86mm]{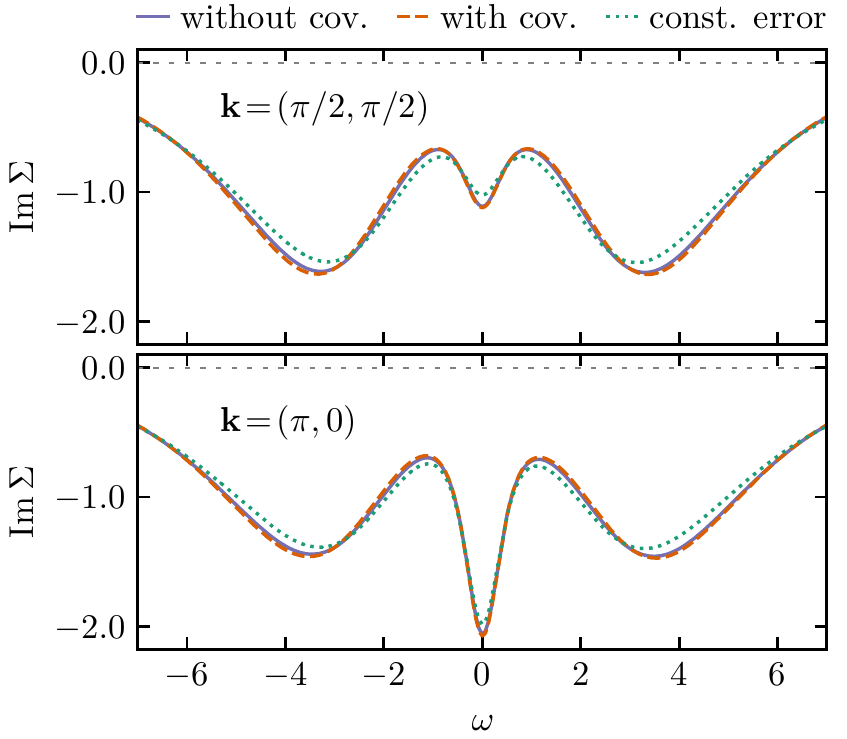}
    \caption{\label{fig:analytic-continuation}
      Imaginary part of the self-energy on the real-frequency axis at different momenta
      $\vect{k}$, comparing the analytic continuation without covariance matrix, with the
      proper covariance matrix, and with a constant error.}
  \end{figure}
  We use the \verb=ana_cont= library\cite{ana-cont,ana-cont-github} to perform analytic
  continuations of the \dga\ self-energy at $\beta=4$. To this end, we have to subtract
  the Hartree energy $U/2$, such that asymptotically also the real part approaches zero.
  Then the standard kernel for fermionic Green's functions can be used\cite{Luttinger1961}
  and we obtain a function $A_\Sigma(\omega)$ which is related to the imaginary part of
  the self-energy by
  \begin{equation}
    A_\Sigma(\omega) = -\frac{1}{\pi} \im \Sigma(\omega + \i0^+).
  \end{equation}
  In \cref{fig:analytic-continuation} we see that in this case the correlations are not
  strong enough to make the analytic continuation instable. However, using a constant
  error rather the one calculated by the jackknife method leads to a slightly different
  result.

\section{Conclusion and discussion}
\label{sec:conclusion}
  We have implemented and studied a jackknife error estimate for typical DMFT and \dga\
  post-processing calculations after a QMC solution of the Anderson impurity problem.
  While mere QMC error bars have been analyzed before (also employing the
  jackknife\cite{DMFT25,SciPostPhys.3.2.013}), the error of the post-processed quantities
  such as the DMFT susceptibilities and the \dga\ self-energies has not been
  systematically studied. Here, the QMC statistical error is propagated through non-linear
  equations, namely the Bethe--Salpeter equation. In such a situation the jackknife method
  is the method of choice, and we make our routines that have been tested with QMC input
  from w2dynamics\cite{w2dynamics} available at Ref.~\onlinecite{jackknife-gitlab}.

  From a statistical point of view, our study reveals that the different components of the
  two-particle Green's function, which is the CT-HYB QMC output and main jackknife input,
  are uncorrelated. But a binning into $\order(100)$ bins is necessary to remove the
  spurious off-diagonal components of the correlation matrix. The same holds for the DMFT
  susceptibility which is calculated through the Bethe--Salpeter equation. Because of the
  vanishing off-diagonal elements of the correlation (or covariance) matrix, an analytical
  continuation without covariance is possible.

  The \dga\ self-energy on the other hand has a non-diagonal correlation matrix. Its
  calculation consists of the Bethe--Salpeter equation, as in DMFT, and additionally the
  Schwinger--Dyson equation. We conclude that the latter leads to the correlations between
  the \dga\ self-energy at different frequencies. This is also quite intuitive since the
  same (bosonic) frequency component of the susceptibility contributes to the \dga\
  self-energy at different fermionic frequencies. However, we have shown that the results
  of analytic continuation of the self-energy are hardly influenced by this. We thus
  conclude that the correlations introduced by two-particle corrections are still small
  enough to allow for a reliable interpretation of the results. Still one should keep in
  mind that using the actual jackknife error, as opposed to a constant, does have a small
  influence on the analytic continuation.

  We have further studied the standard error of the mean (SEM) and observe that it hardly
  depends on the number of bins $n_b$ into which the total number of measurements $n_t$ is
  divided. However as a matter of course the SEM depends strongly on $n_t$. Depending on
  the physical quantity studied and the required accuracy, a total number of measurements
  $n_t$ of $\order(10^7)$ to $\order(10^9)$ is needed. Here, the error bars of the DMFT
  susceptibilities are somewhat larger than for the \dga\ self-energy. The error also
  increases with decreasing temperature or larger interval $[0,\beta]$, as this interval
  is sampled less accurately if $n_t$ is kept fixed.

  Our paper has shown that the jackknife method is a valuable tool for calculating DMFT
  and \dga\ error bars. Beyond the present paper, the statistical error of the
  one-particle Green's function can be included. However, with the use of symmetric
  improved estimators the errors of the one-particle Green's function are practically
  non-existent. A further extension would be to consider the error of the DMFT
  self-consistency loop itself by drawing bins from statistically independent DMFT
  solutions. The latter can be achieved either by completely independent DMFT calculations
  or by determining the auto-correlation time of the DMFT loop and adjusting the binning
  to it. Our approach can be combined with various other methods to reduce the Monte Carlo
  noise\cite{PhysRevB.85.205106,Kaufmann2019,Gunacker2016} or the cutoff error of
  the frequency box\cite{wentzell2016highfrequency,Kaufmann2017,%
  PhysRevB.99.041115,PhysRevB.99.235106,PhysRevB.101.035110} as well as with
  compactifications of the vertex\cite{PhysRevB.84.075145,PhysRevB.97.205111,%
  SciPostPhys.8.1.012}. We hope that our paper will spread the seed for a proper error
  estimate in future DMFT calculations and diagrammatic extensions thereof.
  
\vspace*{2.3mm}
\section*{Acknowledgments}
\label{sec:acknowledgments}
  The present research was supported by the Austrian Science Fund (FWF) through the
  Doctoral School W1243 Building Solids for Function (P. K.) as well as projects P32044
  and P30997. Calculations were done on the Vienna Scientific Cluster (VSC).
\bibliography{Paper,main}{}
\end{document}